\documentclass[aps,twocolumn,nofootinbib]{revtex4}
\usepackage[latin1]{inputenc}
\usepackage{epsfig}
\newcommand{\beq}{\begin{equation}}
\newcommand{\eeq}{\end{equation}}
\newcommand{\la}{\langle}
\newcommand{\ra}{\rangle}

\begin{document}

\title{Stochastic thermodynamics of opinion dynamics}

\author{
Tânia Tomé, Carlos E. Fiore and Mário J. de Oliveira}
\affiliation{Universidade de São Paulo, Instituto de Física,
Rua do Matão, 1371, 05508-090 São Paulo, SP, Brazil}

\begin{abstract}

We show that models of opinion formation and dissemination
in a community of individuals can be framed within stochastic
thermodynamics from which we can build a nonequilibrium
thermodynamics of opinion dynamics. This is accomplished by
decomposing the original transition rate that defines an
opinion model into two or more transition rates, each
representing the contact with heat reservoirs at different
temperatures, and postulating an energy function. As the
temperatures are distinct, heat fluxes are present even
at the stationary state and
linked to the production of entropy, the fundamental 
quantity that characterizes
nonequilibrium states. We apply the present framework to
a generic-vote model including the majority-vote model
in a square lattice and in a cubic lattice. The fluxes and
the rate of entropy production are calculated by
numerical simulation and by the use of a pair approximation.

\end{abstract}

\maketitle

\section{Introduction}

Opinion dynamics \cite{helbing1995,castellano2009} 
deals with the time evolution of the number
of individuals in each of the various groups that a community
is divided by reason of the opinion of the individuals
concerning a given subject. The opinion of an individual
changes with time under the influence of other individuals
or by the influence of an external agent. The repeated action
of these influences leads to a collective behavior of the
individuals in relation to the opinions, which is the
phenomenon to be explained by the opinion models. The
modeling can be accomplished by expressing these influences
in terms of rules that govern the opinion dynamics.

There are various possibilities of setting up opinion
models according to the way one represents the opinion
\cite{martins2008,jedrzejewski2019}. In view of the
contingent character of the influences, we consider
models with dynamics rules that have a stochastic nature. 
A way of representing the stochastic nature of the opinion
dynamics is to consider it to be a continuous time
Markovian process \cite{liggett1985,liggett1999,tome2015L}
defined on a {\it space of opinions}.
This means to say that opinion models are defined once
we are given the rates of the transition from an opinion
state of the whole community to other possible opinion
states. 

Here we focus on models with a discrete space of
opinions in which individuals, or agents,
are located in space at sites whose collection
forms a lattice of sites. To each individual, or to each
site, one associates an {\it opinion variable}, understood
as a random variable, that takes certain discrete numerical
values, each corresponding to a certain opinion
regarding the issue being discussed. The equation
that governs the time evolution of the opinion probability
is the master equation which is set up from the transition
rates.

Many models fall within this framework.
We mention the voter model
\cite{clifford1973,holley1975} in which at each time
step an individual
takes the opinion of one of its neighbors chosen at
random. The noise voter model, or linear Glauber
model, which we call simply linear model,
is a modification of the voter model such that
the individual takes the opinion of the chosen neighbor
with a certain probability and the opposite opinion with the
complementary probability 
\cite{scheucher1988,krapivsky1992,granovsky1995,oliveira2003}.
The original voter model was generalized to the case where
the individual takes the opinion of two or more neighbors
chosen randomly as long as they have a common opinion
\cite{castellano2009a,nyczka2012}. In the Sznajd model 
a pair of neighboring individuals with the same opinions 
convince all its neighbors to their opinion
\cite{sznajd2000}. The voter model was generalized to the
case in which each individual adopts three opinions, for
instance, leftist, centrist and rightist \cite{vazquez2003}.

The majority-vote model \cite{oliveira1992} is based on a
majority rule. An individual takes the opinion of the
majority of its neighbor with a certain probability and
the opposite opinion with the complementary probability.
If there is a tie, the individual takes either opinion
with equal probability. This model displays a critical
phase transition from a disordered to an ordered state
that belongs to the universality class of the Ising model,
in spite of being a non-equilibrium model in the
sense that its transition rate lacks detailed balance.
It was originally defined in the square lattice but
was applied to other regular lattices
\cite{acunalara2012,acunalara2014,sastre2016,chatelain2022}
as well as to small world lattices \cite{campos2003},
to random graphs \cite{pereira2005,lima2005}, and
to complete graphs \cite{fronczak2017}.
When inertia is incorporated to the majority-vote model,
it exhibits a discontinuous phase transition 
\cite{chen2017,harunari2017,encinas2018}.
A version of the majority-vote model with three states
was also conceived \cite{brunstein1999}.

A generic-vote vote model with two states has been
considered with the restriction that the change of opinion
of an individual depends only on the sum of the opinions
in a neighborhood \cite{oliveira1993}. In a square
lattice this model has two parameters. Depending
on the relation between these parameters, the
generic-vote model reduces to
the voter model, the linear model and the majority model.
In general the transition rates do not obey detailed
balance and the generic-vote model is a nonequilibrium
model. However, for a special relation between the two
parameters, it reduces to the Glauber model
\cite{glauber1963,suzuki1968}.

The Glauber model is distinct from the other models
described above in the sense that its transition rate
obeys detailed balance from which follows that the
stationary probability distribution is the Gibbs
probability distribution, which is proportional to 
$e^{-\beta E}$, where $E$ is the energy function.
It describes a system in thermodynamic equilibrium
at a temperature inversely proportional to $\beta$.
For the Glauber model, $E$ is the Ising energy
function and the model defined by the equilibrium Gibbs 
probability distribution is known as the Ising model
and can be interpreted as an opinion model with two
states. In fact, such an interpretation was put
forward as a sociological model and the two
ordered states displayed by the Ising model were
interpreted as the polarizations of opinions found
in society \cite{weidlich1971}.

A similar interpretation was given to the Ising model
with an approach that was supplemented by a thermodynamic
basis \cite{galam1982,galam1991}. Accordingly, the
free energy associated to the Ising model was
interpreted as the dissatisfaction function and the
principle of minimum free energy as a principle of
minimum dissatisfaction. The energy function of the
Ising model was interpreted as measure of the degree
of convergence or divergence, that is, of agreement
or conflict. Although these interpretations are
appealing, they refer to equilibrium thermodynamics
and cannot be extended to include the models presented
above, which cannot be found in thermodynamic
equilibrium as they do not obey detailed balance.

The statistical mechanics that has been used as the
framework to opinion
dynamics \cite{castellano2009}, which encompasses the
master equation as the central evolution equation,
emphasizes the probabilistic aspects but lacks
a thermodynamics perspective as it makes no reference
to energy nor to entropy, the two main concepts of
thermodynamics, and much less to their relationship.

To circumvent these problems, we propose a
framework to the models of opinion dynamics based
on stochastic thermodynamics \cite{tome2010,
esposito2012,seifert2012,vandebroeck2013,tome2015},
which is capable of describing systems out of
thermodynamics equilibrium as well as system in
thermodynamic equilibrium. The stochastic thermodynamics
is appropriate for systems described by a master equation
and can be understood as an enlargement of the statistical
mechanics for systems out of thermodynamic equilibrium.
This approach provides in addition an expression for
the production of entropy, a keystone quantity that
characterizes systems out of thermodynamic equilibrium.

The central and crucial idea that we use in the analysis of
the models for opinion dynamics rests on the decomposition
of the transition rate into several mutually exclusive
transition rates. Each of these transition rates describes
the contact with a heat reservoir at a certain temperature.
The whole transition rate describes thus the contact with
various heat reservoirs at distinct temperatures. As the
temperatures are different, the transition rate does not
obey detailed balance, and as a consequence,
in the stationary state the
system will not be found in thermodynamic equilibrium.
It is worth pointing out that
the idea of using heat reservoir at distinct temperatures
is not new and has been used before. However, the reference
to a heat reservoir was usually nominal and lacks the
crucial Clausius relation between entropy, heat and 
reservoir temperature.

\section{Decomposition of the transition rates}

We consider a community of individuals, each holding an
opinion concerning a particular issue. The individuals
are immobile and are located at the sites of a lattice.
To properly describe the opinion of the whole community,
we attach to each site an opinion variable $\sigma_i$
that takes one of several possible numerical values, each
associated to an opinion concerning the issue at hand.  

The {\it opinion state}, which is the collection of the
opinions of the individuals, is denoted by $\sigma$.
As time goes by, the opinion state changes according
to a stochastic process. The rate at which the state
changes from $\sigma$ to $\sigma'$ is denoted by
$W(\sigma',\sigma$), which is the central quantity that
defined a model of opinion dynamics. The probability
$P(\sigma,t)$ of finding the whole community in a given
opinion $\sigma$ at time $t$ obeys the master equation
\beq
\frac{d}{dt}P(\sigma) = \sum_{\sigma'} \{W(\sigma,\sigma')
P(\sigma') - W(\sigma',\sigma)P(\sigma)\}.
\label{10}
\eeq

The transition rate $W$ is decomposed into a certain
number of transition rates $W_\ell$, 
\beq
W(\sigma,\sigma') = \sum_\ell W_\ell(\sigma,\sigma'),
\label{11}
\eeq
each holding the following property. Given $\sigma$
and $\sigma'$, and if $W_k(\sigma,\sigma')$ is
nonzero then $W_\ell(\sigma,\sigma')$ will vanish 
necessarily for $\ell\neq k$.
The possible pairs of configuration $(\sigma,\sigma')$
are partitioned into a certain number of mutually
exclusive subsets, and one associates to each subset
$\ell$ a transition rate $W_\ell(\sigma,\sigma)$ meaning 
that this transition vanishes if $(\sigma,\sigma')$
does not belong to the subset $\ell$.

Another property of the rate $W_\ell$ is its
association to an energy function. To express this
property one starts by postulating an energy function
$E(\sigma)$, which may be called the {\it opinion
function}. Then the transition rate is set up in
such a way that
\beq
\frac{W_\ell(\sigma',\sigma)}{W_\ell(\sigma,\sigma')}
= e^{-[E(\sigma')-E(\sigma)]/\theta_\ell},
\label{12}
\eeq
where $\theta_\ell$ is a parameter.
We remark that this property is possible if the process
described by the rate $W_\ell$ has it reverse, or
in other terms, if $W_\ell(\sigma',\sigma)$
is nonzero so is $W_\ell(\sigma,\sigma')$.

Let us suppose that the whole transition rate
$W$ is composed by just one transition of the
type (\ref{12}). Then it will hold the property
\beq
\frac{W(\sigma',\sigma)}{W(\sigma,\sigma')}
= e^{-[E(\sigma')-E(\sigma)]/\theta}.
\label{13}
\eeq
If we define the probability distribution 
\beq
P_0(\sigma) = \frac1Z e^{-E(\sigma)/\theta},
\label{14}
\eeq
we see that $P_0$ is the stationary solution of the
master equation, which can be verified by the use of
the property (\ref{13}) which is the detailed balance
condition. The distribution (\ref{14}) is the Gibbs
distribution which describes a system in
thermodynamic equilibrium at a temperature $\theta$.

The whole transition rate $W$ that fulfills 
the detailed balance condition (\ref{13}) can
also be understood as the one appropriate
to describe a system in contact with a heat
reservoir at a temperature $\theta$.
When $W$ is a sum of terms $W_\ell$, the
detailed balance condition is not satisfied.
At first sight the property (\ref{12}) seems to
be the detailed balance condition but it is not
because $\theta_\ell$ in (\ref{12}) is distinct
for distinct $\ell$. 
However, we may interpret $W$ as describing
a system in contact with several heat reservoir
each being at a given temperature $\theta_\ell$.
In the long run, when the steady state is reached,
the system will not be found in the equilibrium state
since the detailed balance is not satisfied, unless all
temperatures are the same.

\section{Stochastic thermodynamics}

\subsection{General}

The time derivative of the average energy
$U=\la E(\sigma)\ra$ is obtained by multiplying the
master equation (\ref{10}) by $E(\sigma)$ and
summing in $\sigma$. The result is
\beq
\frac{dU}{dt} = \Phi,
\label{20a}
\eeq
\beq
\Phi = \sum_\ell \Phi_\ell,
\label{20b}
\eeq
\beq
\Phi_\ell = \sum_{\sigma\sigma'}
[E(\sigma') - E(\sigma)]W_\ell(\sigma',\sigma)P(\sigma).
\label{15}
\eeq
The quantity $\Phi_\ell$ is understood as the flux of
energy {\it from} the $\ell$-th heat reservoir {\it to}
the system, and $\Phi$ is the total flux of energy
from the reservoirs to the system.

The entropy of the system is defined by
\beq
S = - \sum_\sigma P(\sigma)\ln P(\sigma),
\eeq
and depends on time. Its time derivative can be
written by the use of the master equation (\ref{10})
in the following form
\beq
\frac{dS}{dt} = \Pi - \Psi,
\label{22a}
\eeq
\beq
\Pi = \sum_\ell \Pi_\ell,
\qquad
\Psi = \sum_\ell \Psi_\ell,
\label{22b}
\eeq
where
\[
\Pi_\ell = \frac12\sum_{\sigma\sigma'}
\{W_\ell(\sigma,\sigma')P(\sigma'
- W_\ell(\sigma',\sigma)P(\sigma)\}
\times
\]
\beq
\times
\ln \frac{W_\ell(\sigma,\sigma')P(\sigma')}
{W_\ell(\sigma',\sigma)P(\sigma)},
\eeq
\beq
\Psi_\ell = \sum_{\sigma\sigma'}
W_\ell(\sigma',\sigma)P(\sigma)
\ln \frac{W_\ell(\sigma',\sigma)}{W_\ell(\sigma,\sigma')}.
\eeq
The quantity $\Pi_\ell$ is the rate of the entropy
production due to process $\ell$ and is a {\it nonnegative}
quantity, and $\Pi$ is the total rate of entropy production.
The quantity $\Psi_\ell$ is interpreted as the flux of
entropy {\it from} the system {\it to} the reservoir
$\ell$, and $\Psi$ is the total flux of entropy from
the system to the reservoirs. 

Next we use the property (\ref{12}) to 
write the flux of entropy $\Psi_\ell$ in the form
\beq
\Psi_\ell = -\frac1{\theta_\ell}\sum_{\sigma\sigma'}
W_\ell(\sigma',\sigma)P(\sigma)
[E(\sigma')-E(\sigma)].
\eeq
Comparing this expression with (\ref{15}), we reach 
the following relation between the flux of entropy
and the flux of energy related to the reservoir $\ell$,
\beq
\Psi_\ell = -\frac{\Phi_\ell}{\theta_\ell}.
\label{25}
\eeq

It is worth making the following comment
concerning the stationary state. In this state
$dU/dt=0$ and $dS/dt=0$, from which follows that
\beq
\sum_\ell \Phi_\ell = 0,
\eeq
and hence
\beq
\Pi = -\sum_\ell \frac{\Phi_\ell}{\theta_\ell}.
\label{16}
\eeq
Although the sum of all fluxes of energy vanishes in
the stationary state it does not mean that the sum
on the right-hand side of (\ref{16}) will vanish.
This happens because the temperature are distinct from
each other. Thus there is a continuous production of
entropy and the system is out of thermodynamic
equilibrium. The equilibrium will be reached if
all temperatures are the same in which case the
right-hand side of (\ref{16}) vanishes. In this
case the production of entropy vanishes and the
system will be found in equilibrium.

A special type of transition rate could also be included
in the decomposition (\ref{11}) of $W(\sigma,\sigma')$.
This type of transition rate, which we label by $\ell=0$
is the one such that 
$W_0(\sigma,\sigma')=W_0(\sigma',\sigma)$.
The corresponding contribution to the
production of entropy is
\beq
\Pi_0 = \frac12\sum_{\sigma\sigma'}
W_0(\sigma,\sigma')\{P(\sigma')
- P(\sigma)\}\ln \frac{P(\sigma')}{P(\sigma)},
\eeq
but the corresponding contribution to the flux of entropy
vanishes. We assume that the variation in energy associated
to this special type of transition rate vanishes,
$E(\sigma')=E(\sigma)$, which is consistent with the
relation (\ref{13}). Thus, the contribution to the
flux of energy also vanishes. 

The relation (\ref{25}), which is a
fundamental result coming out of the present approach, 
is identified as the essential and central
relation of thermodynamics introduced by Clausius. However,
the Clausius relation connects the entropy to the heat
flux and not to the energy flux. To conform with Clausius
we proceed as follows.
The energy function $E(\sigma)$ that we have postulated
and which appears in the definition (\ref{12}) of the
transition rates $W_\ell$ is replaced by the function 
$H(\sigma)$, which is $E(\sigma)$, understood as the 
internal energy, subtracted from the potential energy
$L(\sigma)$ due to external forces, that is, $H=E-L$. 

The time evolution of $U=\la E\ra$ is given by 
equations (\ref{20a}), (\ref{20b}), and (\ref{15}),
and that of $\la L\ra$ is given by
\beq
\frac{d\la L\ra}{dt} = \Phi^{\rm ext}
= \sum_\ell \Phi_\ell^{\rm ext},
\eeq
\beq
\Phi_\ell^{\rm ext} = \sum_{\sigma\sigma'}
[L(\sigma') - L(\sigma)]W_\ell(\sigma',\sigma)P(\sigma).
\label{15a}
\eeq
Using the same reasoning above we conclude that
the entropy flux $\Psi_\ell$ from the system to the
reservoir $\ell$ is now given by
\beq
\Psi_\ell = -\frac{\Phi_\ell^{\rm q}}{\theta_\ell},
\eeq
where 
\beq
\Phi_\ell^{\rm q} = \Phi_\ell-\Phi_\ell^{\rm ext}.
\eeq
Writing $\Phi_\ell=\Phi_\ell^{\rm q} + \Phi_\ell^{\rm ext}$
we see that this equation can be understood as
the conservation of energy and $\Phi_\ell^{\rm q}$
as the flux of heat. The first term is the variation
of energy of the system and the last is the
variation of the external potential per unit time,
or power. From now on we treat the cases such that
$L$ can be set to zero and the flux of heat is
identified as the flux of energy.

\subsection{One-site transitions}

We consider here models such that at each time step just
one individual changes its opinion. The individuals are
located at the sites of a lattice and we suppose that
each individual holds an opinion pro or against a
particular issue. The opinion variable $\sigma_i$ then
takes two values which we choose to be $+1$ or $-1$
according to whether the individual at the site $i$
of the lattice is pro or against the issue at hand,
respectively. Thus the dynamics is defined by the
transition rate $w_i(\sigma)$ that the individual
at $i$ changes its opinion from the present to
the opposite opinion, that is, from $\sigma_i$
to $-\sigma_i$, and the master equation reads
\beq
\frac{d}{dt}P(\sigma) = \sum_i \{w_i(\sigma^i)P(\sigma^i)
- w_i(\sigma)P(\sigma)\},
\label{45}
\eeq
where $\sigma^i$ stands for the state obtained from
$\sigma$ by changing $\sigma_i$ to $-\sigma_i$. The
one-site transition rate is decomposed in the form
\beq
w_i(\sigma) = \sum_\ell w_{\ell i}(\sigma),
\eeq
where $w_{\ell i}$ holds the property
\beq
\frac{w_{\ell i}(\sigma)}{w_{\ell i}(\sigma^i)}
= e^{-[E(\sigma^i)-E(\sigma)]/\theta_\ell},
\label{23}
\eeq
where $E(\sigma)$ is the postulated energy.
The transition rates $w_{\ell i}(\sigma)$ holds
the property mentioned just below the equation (\ref{11}).
That is, if $\sigma$ is a state such that $w_{ki}(\sigma)$
is nonzero than $w_{\ell i}(\sigma)$ will vanish
for $\ell\neq k$.

The time variation of the average energy $U$
is given by (\ref{20a}) and (\ref{20b}),
where now the energy flux from the reservoir $\ell$ reads
\beq
\Phi_\ell = \sum_\sigma \sum_i [E(\sigma^i) - E(\sigma)]
w_{\ell i}(\sigma)P(\sigma),
\eeq
which can be written as an average
\beq
\Phi_\ell = 
\sum_i \langle [E(\sigma^i)
- E(\sigma)] w_{\ell i}(\sigma)\rangle,
\label{4}
\eeq
and can thus be obtained from numerical simulations.

The time variation of the entropy is given by 
(\ref{22a}) and (\ref{22b}),
where now the rate of entropy production due to the process
$\ell$ is
\[
\Pi_\ell = \frac12\sum_\sigma \sum_i \{w_{\ell i}(\sigma^i)
P(\sigma^i) - w_{\ell i}(\sigma)P(\sigma)\}\times
\]
\beq
\times\ln\frac{w_{\ell i}(\sigma^i)P(\sigma^i)}
{w_{\ell i}(\sigma)P(\sigma)},
\eeq
and is a nonnegative quantity, and the flux of entropy
$\Psi_\ell$ from the system to the reservoir $\ell$ is
\beq
\Psi_\ell = \sum_\sigma \sum_i
w_{\ell i}(\sigma)P(\sigma) \ln\frac{w_{\ell i}(\sigma)}
{w_{\ell i}(\sigma^i)}.
\label{53}
\eeq
Using the relation (\ref{23}) we reach again the
result (\ref{25}), namely,
$\Psi_\ell=-\Phi_\ell/\theta_\ell$. 
We remark that the $\Psi_\ell$ given by (\ref{53})
can be written as the following average
\beq
\Psi_\ell = \sum_i\left\la
w_{\ell i}(\sigma) \ln\frac{w_{\ell i}(\sigma)}
{w_{\ell i}(\sigma^i)}\right\ra.
\label{53a}
\eeq

\section{Contact with one heat reservoir}

A transition rate $w_i$ that leads to the Gibbs
equilibrium probability distribution
\beq
P_0(\sigma) = \frac1Z e^{E(\sigma)/\theta}
\eeq
can be constructed by the use of the detailed balance
condition
\beq
\frac{w_i(\sigma)}{w_i(\sigma^i)} = \frac{P_0(\sigma^i)}{P_0(\sigma)}
= e^{-[E(\sigma^i)-E(\sigma)]/\theta}.
\eeq
The most general form of $w_i$ satisfying the detailed
balance condition is
\beq
w_i(\sigma) = k_i(\sigma)e^{E(\sigma)/\theta},
\eeq
where $k_i(\sigma^i)=k_i(\sigma)$, that is, $k_i(\sigma)$
does not depend on $\sigma_i$.
If we define $\alpha_i(\sigma_i)$ by
\beq
\alpha_i(\sigma) = k_i(\sigma)[e^{E(\sigma^i)/\theta}+e^{E(\sigma)/\theta}],
\eeq
we see that the transition rate can be written in the form
\beq
w_i(\sigma) = \frac12\alpha_i(\sigma)
[1-\tanh\frac{E(\sigma^i)-E(\sigma)}{2\theta}],
\eeq
where $\alpha_i(\sigma)$, as happens to $k_i(\sigma)$,
does not depend on $\sigma_i$,
that is, $\alpha_i(\sigma^i)=\alpha_i(\sigma)$,
which is also a general form of $w_i(\sigma)$
satisfying the detailed balance condition.
Since this transition leads to the equilibrium Gibbs 
distribution, we may interpret it as describing the
contact with a heat reservoir at a temperature $\theta$.

If the energy function is
\beq
E(\sigma) = -\sum_{(ij)} \sigma_i\sigma_j,
\label{30}
\eeq
where the summation is over all nearest neighbor pairs
of sites, then 
\beq
w_i(\sigma) = \frac12\alpha_i(\sigma)
[1-\sigma_i\tanh\frac{1}{\theta}
\sum_\delta\sigma_{i+\delta}],
\label{31}
\eeq
where the summation is over the nearest
neighbor sites of site $i$.

When $\alpha_i$ does not depend on $\sigma$, that is, 
when it is a constant, 
\beq
w_i(\sigma) = \frac\alpha2
[1-\sigma_i\tanh\frac{1}{\theta}
\sum_\delta\sigma_{i+\delta}],
\label{31c}
\eeq
and it is called Glauber transition rate because Glauber
used it to describe the dynamics of the one-dimensional
Ising model. In this case the expression (\ref{31c})
can be simplified and reads
\beq
w_i(\sigma) = \frac\alpha2
[1-\frac\gamma2 \sigma_i (\sigma_{i-1}+\sigma_{i+1})],
\label{31d}
\eeq
where $\gamma=\tanh 2/\theta$. In fact, this is the
originalexpression used by Glauber in 1963
\cite{glauber1963}. That given by (\ref{31c})
was introduced later on by Suzuki and Kubo
in 1968 \cite{suzuki1968}.

\section{Generic-vote model}

\subsection{General}

We apply the results of the previous section to the
generic-vote model introduced in reference
\cite{oliveira1993}. In this model the change of opinion
of an individual depends only on the sum of the opinions
of the nearest neighbor individuals. We consider here a
square lattice in which case the number
of nearest neighbors is four, and the possible values
of the sum are $4,2,0,-2,-4$. The rates at which the 
opinion changes are given in table \ref{rates}.

We will consider the transition rate with reference
to a certain site labeled 0.
In addition, the four nearest neighbor of site 0 are
labeled 1, 2, 3, and 4, which are respectively 
on the right, above, left and below the site zero.
The transition rate related to the site zero is
assumed to be of the form
\beq
w = \frac12(1-\sigma_0 f),
\label{42}
\eeq
where $f$ is a function of the opinion variables
associated to the four nearest neighbor of site 0.
Notice that we are dropping the site index of $w$ and $f$.
For the generic-vote model $f$ is given by
\cite{oliveira1993}
\beq
f = s(a+b\sigma_1\sigma_2\sigma_3\sigma_4),
\label{33}
\eeq
\beq
s = \sigma_1+\sigma_2+\sigma_3+\sigma_4,
\eeq
where $a$ and $b$ are parameters related to $p$ and $q$ by
\beq
a=\frac18(p+2q), \qquad b=\frac18(p-2q),
\eeq
or $p=4(a+b)$ and $q=2(a-b)$.

Replacing these relations in (\ref{33}), we
write 
\beq
f = q \,\xi_1 + p \,\xi_2,
\label{34}
\eeq
where
\beq
\xi_1 = \frac{s}4(1-\sigma_1\sigma_2\sigma_3\sigma_4),
\label{50b}
\eeq
\beq
\xi_2 = \frac{s}8(1+\sigma_1\sigma_2\sigma_3\sigma_4).
\label{50a}
\eeq
Notice $\xi_1$ takes the values $1$ and $-1$ for $s$ equal
to $2$ and $-2$, and the value zero otherwise; and 
that $\xi_2$ takes the values $1$ and $-1$ for $s$ equal
to $4$ and $-4$, and the value zero otherwise. 

The  transition rate $w$ is invariant
under the inversion of all variables $\sigma_i$, which
amounts to say that $f$ changes sign under the
inversion transformation. The expression (\ref{33})
is the most generic form for $f$ obeying this invariance
\cite{oliveira1993}. As shown in table \ref{rates},
the transition rate $w$ for the generic-vote model
depends only on the four nearest neighbor
variables only through their sum $s$.

\begin{table}
\caption{Transition rates of the generic-vote model in
a square lattice. The transition rates depend on the
possible values $4,2,0,-2,-4$ of the sum of the opinion 
variables of the neighboring individuals.}
\begin{tabular}{|c|c|c|c|c|c|}
\hline
$\sigma_0\to-\sigma_0$ & 4   & 2   & 0    & -2  & -4 \\
\hline
$1\to -1$  & $\frac12(1-p)$ & $\frac12(1-q)$ &
$\frac12$  & $\frac12(1+q)$ & $\frac12(1+p)$ \\
$-1\to 1$  & $\frac12(1+p)$ & $\frac12(1+q)$ &
$\frac12$  & $\frac12(1-q)$ & $\frac12(1-p)$ \\
\hline
\end{tabular}
\label{rates}
\end{table}

The decomposition of the transition rate $w$ is carried
out by postulating an energy function, which we
choose to be that given by (\ref{30}). The
transition rate is written as the sum of three transition
rates,
\beq
w =  g_0 + g_1 + g_2,
\label{37}
\eeq
two of them being of the type (\ref{31}),
\beq
g_1 = \frac12\alpha_1[1-\sigma_0\tanh\frac{s}{\theta_1}],
\label{31b}
\eeq
\beq
g_2 = \frac12\alpha_2]1-\sigma_0\tanh\frac{s}{\theta_2}],
\label{31a}
\eeq
which describe the contact with heat reservoirs at
temperatures $\theta_1$ and $\theta_2$, respectively.
The third component $g_0$ is chosen not depend on
$\sigma_0$ so that $g_0(\sigma^i)=g_0(\sigma)$.
We write $g_0 = \alpha_0/2$.

The first process associated to the transition rate
$g_1$, which describes the contact with the
first reservoir at temperature $\theta_1$, occurs when
$s=\pm2$; and the second process associated to the
transition rate $g_2$, which describes the
contact with the second reservoir at temperature
$\theta_2$, occurs when $s=\pm4$; and we assume that
the process labeled zero occurs when $s=0$.
These assumptions are accomplished by setting 
$\alpha_1$ equal to $1$ for $s\pm2$, and zero otherwise;
$\alpha_2$ equal to $1$ for $s\pm4$, and zero otherwise;
and $\alpha_0$ equal to $1$ for $s=0$, and zero otherwise.

If we consider the transition $1\to-1$ and $s=2$,
then $w=g_1=(1-\tanh2/\theta_1)$ which, according to
table \ref{rates}, must be equal to $(1-q)/2$, and 
we conclude that
\beq
q=\tanh\frac{2}{\theta_1}.
\label{36b}
\eeq
Analogously, if we consider the transition $1\to1$ and
$s=4$, then $w=g_2=(1-\tanh4/\theta_2)$ which, according
to table \ref{rates}, must be equal to $(1-p)/2$, and 
we conclude that
\beq
p=\tanh\frac{4}{\theta_2}.
\label{36a}
\eeq

The relations of the parameters $p$ and $q$ with the
temperatures $\theta_1$ and $\theta_2$, given by
(\ref{36b}) and (\ref{36a}), can also be
obtained as follows. We start by writing the following
equivalent forms of $g_1$ and $g_2$,
\beq
g_1 = \frac12\alpha_1 [1-\sigma_0\frac{s}2
\tanh\frac{2}{\theta_1}],
\label{32b}
\eeq
\beq
g_2 = \frac12\alpha_2 [1-\sigma_0 \frac{s}4
\tanh\frac{4}{\theta_2}],
\label{32a}
\eeq
and by replacing them in (\ref{37}). The result for
$w$ is of the form (\ref{42}) with $f$ is given by
\beq
f = \alpha_1\frac{s}2\tanh\frac{2}{\theta_1}
+ \alpha_2\frac{s}4\tanh\frac{4}{\theta_2},
\eeq
where we used the relation
$\alpha_0+\alpha_1+\alpha_2=1$.
Taking into account that
$\alpha_1 s = 2 \xi_1$ and $\alpha_2 s =4 \xi_2$,
we reach the result
\beq
f = \xi_1 \tanh\frac{2}{\theta_1}
+ \xi_2 \tanh\frac{4}{\theta_2},
\eeq
which compared with (\ref{34}) leads us to the
results (\ref{36b}) and (\ref{36a}).

\subsection{Specific models}

\begin{figure}
\epsfig{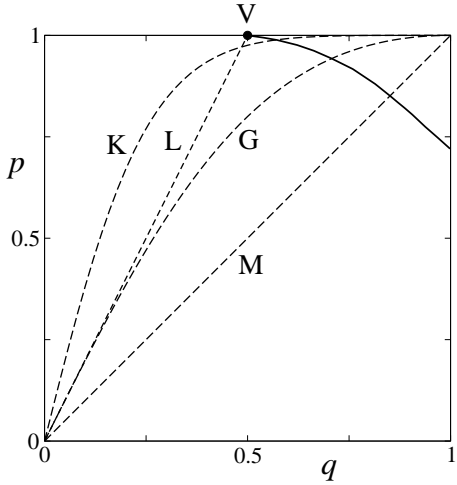}
\caption{Phase diagram of the generic-vote model
in a square lattice in the plane
$q=\tanh2/\theta_1$ and $p=\tanh4/\theta_2$. 
The dashed lines correspond to majority-vote (M) model,
Glauber (G) model, model K, and linear (L) model.
The small full circle (V) corresponds to the voter model. 
Below the line G,
$\theta_2>\theta_1$, and above it $\theta_2<\theta_1$.
Along the line G, $\theta_2=\theta_1$. The solid line
is the critical line separating the ordered state,
above the line, from the disordered state, below the
line. The critical point of the models K, G, and M
occur at $q_c=0.576(2)$ $q_c=\sqrt{2}/2$, and
$q_c=0.850(2)$ \cite{oliveira1992}, respectively.}
\label{diag}
\end{figure}

\begin{figure}
\epsfig{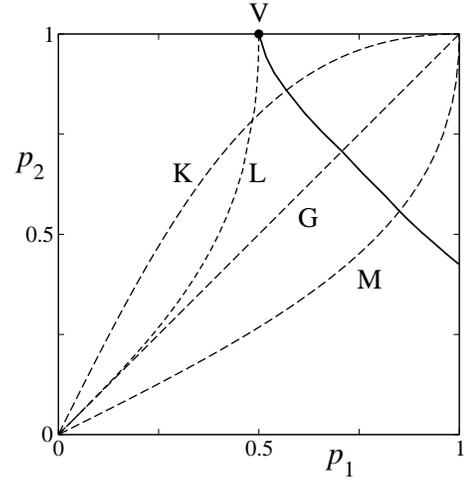}
\caption{Same as figure \ref{diag} but in the
plane $p_1=\tanh2/\theta_1$ and $p_2=\tanh2/\theta_2$.}
\label{diagT}
\end{figure}

Depending on the values of the parameters $q$ and $p$,
several models are particular cases of the generic-vote
model. They correspond to a certain relation between
$q$ and $p$, and thus described by lines in the
diagrams of figures \ref{diag} and \ref{diagT}.
The majority-vote model \cite{oliveira1992} is
defined by the transition rate
\beq
w_{\rm M} = \frac12(1-\gamma\sigma_0 {\cal S}),
\eeq
where $\gamma$ is a parameter and ${\cal S}(s)$
is a function of $s$ that
takes the values $-1,0,+1$, when $s<0$, $s=0$,
$s>0$, respectively.
We see that the transition rate $w$ 
reduces to the majority-vote transition rate when 
$p=q=\gamma$ or equivalently when $a=-3b=3\gamma/8$.
The temperatures are related to $\gamma$ by
\beq
\theta_1 = \frac{4}{\ln(1+\gamma)/(1-\gamma)},
\eeq
and $\theta_2=2\theta_1$.

When $b=0$, the model reduces to the linear model 
\cite{oliveira2003}, defined by the transition rate
\beq
w_{\rm L} = \frac12(1 - a \sigma_0 s).
\eeq
In this case $p=2q=4a$, and
\beq
\theta_1 = \frac{4}{\ln(1+2a)/(1-2a)},
\eeq
\beq
\theta_2 = \frac{8}{\ln(1+4a)/(1-4a)}.
\eeq

If $b=0$ and in addition $a=1/4$, then the model
reduces to the voter model \cite{holley1975},
defined by the transition rate
\beq
w_{\rm V} = \frac12(1 - \frac14 \sigma_0 s).
\eeq
In this case $p=1$, and $q=1/2$, and
\beq
\theta_1 = \frac{4}{\ln 3},
\eeq
and $\theta_2\to0$. The vanishing of $\theta_1$
is a consequence of the fact that when $s=\pm4$,
the reverse transition rate vanishes.

The Glauber transition rate is given by
\beq
w_{\rm G} = \frac12(1-\sigma_0 \tanh \frac{s}{\theta}),
\eeq
and describes the contact of a system with energy
function (\ref{30}) with a reservoir at temperature
$\theta$. Using the relation
\beq
\tanh\frac{s}{\theta} = \xi_1 \tanh\frac{2}{\theta}
+ \xi_2 \tanh\frac{4}{\theta},
\eeq
it follows that the generic-vote transition rate reduces
to the Glauber transition rate if $q=\tanh2/\theta$ and
$p=\tanh4/\theta$, that is,
\beq
\theta_1 = \theta_2 = \theta.
\eeq
From this relation between it follows that $p$ and $q$ 
are connected by
\beq
p = \frac{2q}{1+q^2}.
\eeq

As an example of a model whose fluxes of energy is
the opposite of the majority vote model we
define a model corresponding to a temperature
$\theta_2$ smaller than $\theta_1$. We choose
$\theta_2=\theta_1/2$, and call it model K.
Using (\ref{36b}) and (\ref{36a}), the relation
between $p$ and $q$ for model K is
\beq
p = \frac{4q(1+q^2)}{1+6q^2+q^4}.
\eeq

\section{Fluxes of energy and entropy production}

\subsection{Square lattice}

There are two quantities that interest us here which
are the fluxes of energy and the rate of entropy
production in the stationary state. From the formula
(\ref{4}), we see that there is no flux of energy due
to the zero process because in this case the energy
$E$ remains the same. The energy fluxes per site due
to the first and second processes are
\beq
\phi_1 = 2\la \sigma_0 s g_1\ra,
\qquad
\phi_2 = 2\la \sigma_0 s g_2\ra,
\label{41}
\eeq
because $E(\sigma^0)-E(\sigma)=2\sigma_0 s$.
In the stationary state the rate of entropy production
per site is given by
\beq
{\cal P} = -\frac{\phi_1}{\theta_1}
- \frac{\phi_2}{\theta_2},
\eeq
and taking into account that $\phi_1+\phi_2=0$,
we may write
\beq
{\cal P} = \phi_2\left(\frac{1}{\theta_1}
- \frac{1}{\theta_2}\right).
\label{51}
\eeq

\begin{figure}
\epsfig{file=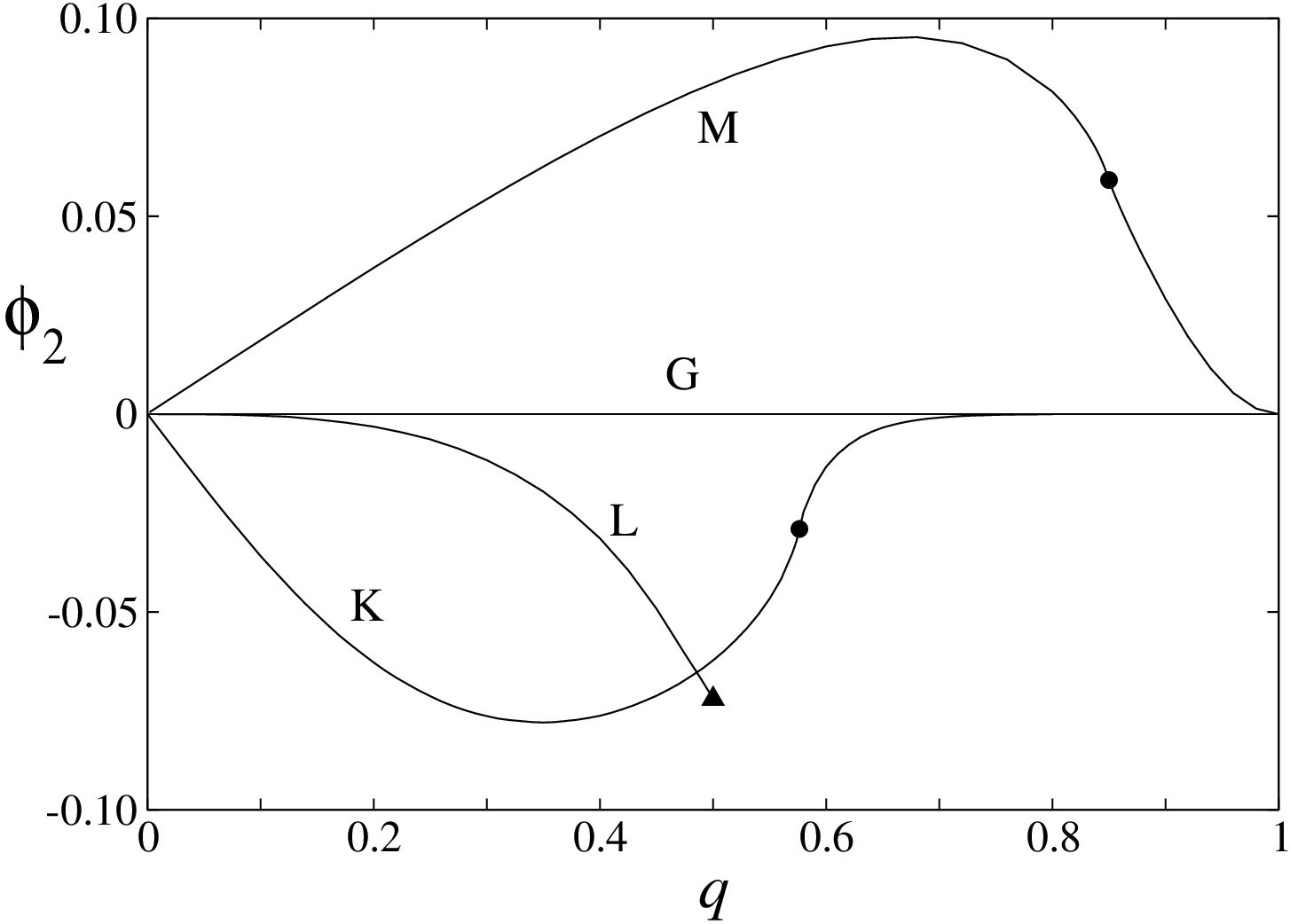,width=8.5cm}
\caption{Flux of energy $\phi_2$ from the second to the
firs reservoir as a function of $q$ for the majority-vote
model (M), for the linear model (L), and for the model K.
For the Glauber model (G) the flux vanishes. The critical
points of models M and K occur at the inflexion point and
are indicated by small full circles. The triangle represents
the flux of energy for the voter model. The results
were obtained by numerical simulations on a square
lattice with $100\times100$ sites.}
\label{fluxsqua}
\end{figure}

The quantity $\phi_2$ represents the flux of energy
that traverses the system from the second to the
first reservoir. Taken into account that ${\cal P}>0$
then if $\theta_2>\theta_1$ then $\phi_2>0$ and energy
flows from the second to the first reservoir.
This happens to the models with parameters $p$ and $q$ 
in the region below the line G of the diagrams of
figures \ref{diag} and \ref{diagT}. Above this line $\theta_2<\theta_1$,
and since ${\cal P}>0$, then $\phi_2<0$ and energy 
flows effectively from the first to the second reservoir
as happens to model K and the linear model.
Along the line G, the energy flux $\phi_2$ vanishes. 

\begin{figure}
\epsfig{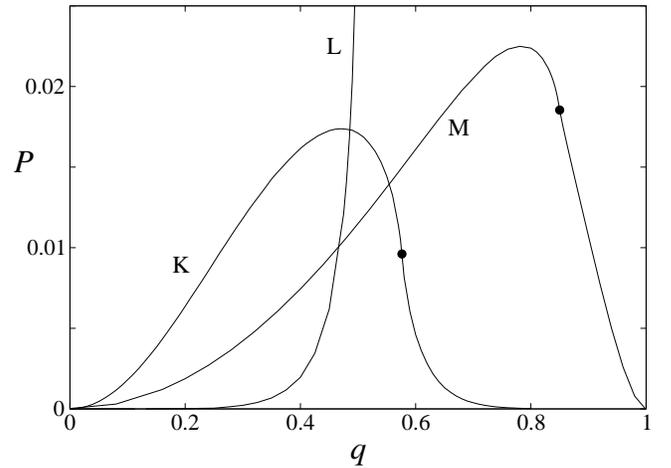}
\caption{Rate of entropy production ${\cal P}$ as
a function of $q$ for the majority-vote model (M),
for the linear model (L), and for the model K.
The critical point of models M and K occur at the
inflexion point and are indicated by small full circles.
The curve L diverges at $q=0.5$. 
The results were obtained from the fluxes shown in
figure \ref{fluxsqua} by the use of formula (\ref{51}).}
\label{prodsqua}
\end{figure}

From equation (\ref{41}), we see that the energy fluxes
are averages and can thus be calculated from numerical
simulations. We have simulated the majority-vote model,
the model K and the linear model in a square lattice
with sizes $100\times100$ sites with periodic boundary
conditions. The fluxes of energy as a function of $q$
are shown if figure \ref{fluxsqua}. It is positive
for the majority-vote model, and negative for the
model K and for linear model as expected.
From the flux of energy, the rate of the production
of entropy is determined from equation (\ref{51})
and is shown in figure \ref{prodsqua}.
The entropy production of the majority-vote model
in a square lattice 
has already been calculated in a direct manner
by the use of equation (\ref{53a}) \cite{crochik2005}.
This equation has also been used to calculate
the production of entropy in other models
and as a means of characterizing nonequilibrium
phase transitions \cite{noa2019}.

The generic-vote model is found to undergo a phase
transition \cite{oliveira1993} from a disordered state
characterized by $\la\sigma_i\ra=0$,
occurring for small values of $q$ and $p$, to an ordered
state characterized by $\la\sigma_i\ra\ne0$,
occurring at higher values of $q$ and $p$, as
shown in figures \ref{diag} and \ref{diagT}. 
When one moves from inside the ordered state region
to the transition line, $\la\sigma_i\ra$ vanishes
continuously. The criticality behavior of the model is
reflected as a singularity in $\phi$ and ${\cal P}$
at the inflexion points of these quantities
which occur at $q_c=0.850(2)$ for the majority-vote
model \cite{oliveira1992} and $q_c=0.576(2)$ for the model K.
The linear model becomes critical at the voter
point occurring at $q=0.5$. At this point,
$\phi_2$ is finite but ${\cal P}$ diverges.

It should be pointed out that for finite lattices the
slope at the inflexion point of the fluxes and the
rate of entropy production is finite. As one increases
the size of the lattice, the slope becomes greater
and diverges in the thermodynamic limit, characterizing
a true singularity \cite{crochik2005}. As we shall see
below, the same is true for the model defined in a
cubic lattice.

\subsection{Cubic lattice}

The generic-vote model can also be defined in a cubic 
lattice. In this case it is necessary to introduce 
three heat reservoirs to describe appropriately
the transition rates. The six nearest neighbor sites
to a central site 0 are labeled $1,2,3,4,5,6$ 
and we denote by $s$ the sum
$s=\sigma_1+\sigma_2+\sigma_3+\sigma_4+\sigma_5+\sigma_6$
and by $\theta_1$, $\theta_2$, and $\theta_3$ the
temperatures of the three reservoirs. 
The transition rate is 
\beq
w = g_0 + g_1 + g_2 + g_3,
\eeq
where $g_0=\alpha_0/2$, and
\beq
g_1 = \frac12(\alpha_1-\sigma_0 \xi_1 q),
\eeq
\beq
g_2 = \frac12(\alpha_2-\sigma_0 \xi_2 p),
\eeq
\beq
g_3 = \frac12(\alpha_3-\sigma_0 \xi_3 r),
\eeq
where $q=\tanh2/\theta_1$, $p=\tanh4/\theta_2$,
$r=\tanh6/\theta_3$; and
$\alpha_0=1$ when $s=0$, and zero otherwise;
$\alpha_1=1$ when $s=\pm2$, and zero otherwise; 
$\alpha_2=1$ when $s=\pm4$, and zero otherwise;
and $\alpha_3=1$ when $s=\pm6$, and zero otherwise;
whereas $\xi_1$ takes the values $1$ and $-1$ for $s$
equal to $2$ and $-2$, and zero otherwise;
$\xi_2$ takes the values $1$ and $-1$ for $s$
equal to $4$ and $-4$, and zero otherwise;
and $\xi_3$ takes the values $1$ and $-1$ for $s$
equal to $6$ and $-6$, and zero otherwise.

\begin{figure}
\epsfig{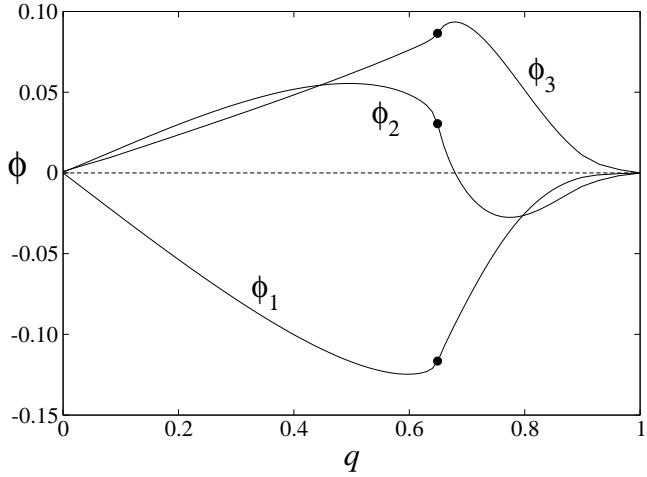}
\caption{Fluxes of energy $\phi_1$, $\phi_2$, and
$\phi_3$ as functions of $q$ for the majority-vote model
in a cubic lattice. The critical transition point occurs
at the inflexion point of the curves and are indicated
by small full circles. The results were obtained by
numerical simulations on a cubic lattice with
$20\times20\times20$ sites. }
\label{fluxcubic}
\end{figure}

The fluxes of energy per site are given by
\beq
\phi_\ell = 2\la\sigma_0 s g_\ell\ra,
\eeq
and the rate of entropy production per site
is given by
\beq
{\cal P} = -\frac{\phi_1}{\theta_1}
- \frac{\phi_2}{\theta_2} - \frac{\phi_3}{\theta_3}.
\label{52}
\eeq

The generic-vote model on the cubic line,
defined by the transition rates above, reduces to the
majority-vote model on the cubic lattice when
$p=q=r$ which implies that the temperature
are related by $\theta_1=\theta_2/2=\theta_3/3$.
We have simulated the majority-vote model on a
cubic lattice with $20\times20\times20$ sites
with periodic boundary conditions.
We determined the three energy fluxes
$\phi_1$, $\phi_2$, and $\phi_3$ in the stationary
state and they are shown in figure \ref{fluxcubic}. 
Within the statistical errors we verified that
$\phi_1+\phi_2+\phi_3=0$. The model displays a
critical phase transition that occurs at the
inflexion points of the fluxes which we
found to be $q_c=0.649(2)$ which is in agreement
with previous calculation on the location of
the critical point, $q_c=0.6474(2)$
\cite{acunalara2012}.

\begin{figure}
\epsfig{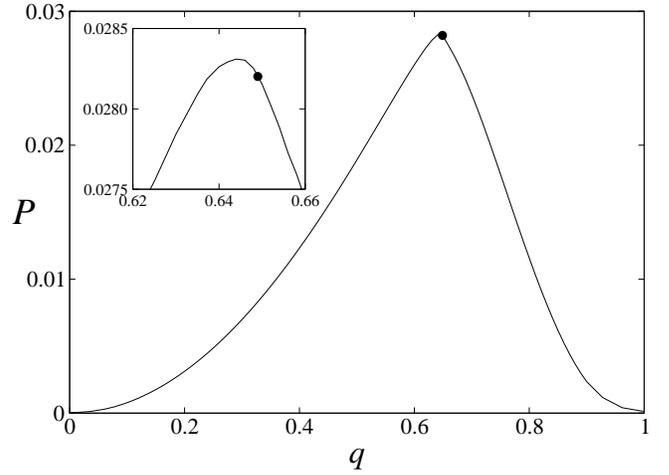}
\caption{Rate of entropy production ${\cal P}$ as a
function of $q$ for the majority-vote model in a cubic
lattice. The critical transition point is indicated by
a small full circle and does not occur at the maximum
of the curve as can be seen in the inset.
The results were obtained from the fluxes shown in
figure \ref{fluxcubic} by the use of formula (\ref{52}).
}
\label{prodcubic}
\end{figure}

We have also determined the rate of entropy
production ${\cal P}$, which is shown in figure
\ref{prodcubic} and we see that it is in fact positive
although the three fluxes do not have the same sign.

\section{Pair approximation}

We solve the master equation (\ref{45}) by means of 
a pair approximation \cite{tome1989}. Denoting by
$P(\sigma_1,\sigma_2,\sigma_3,\sigma_4|\sigma_0)$
the conditional probability associated to the 
four neighboring sites of site zero, the pair
approximation consists in writing
\beq
P(\sigma_1,\sigma_2,\sigma_3,\sigma_4|\sigma_0)
= \prod_i P(\sigma_i|\sigma_0),
\label{46}
\eeq
where
$P(\sigma_i|\sigma_0)=P(\sigma_i,\sigma_0)/P(\sigma_0)$.
The pair probability is parametrized as follows
\beq
P(\sigma_i,\sigma_0) = \frac14[1+m(\sigma_0+\sigma_i)
+ r\sigma_0\sigma_i],
\eeq
where $m=\la \sigma_0\ra=\la\sigma_i\ra$ and
$r=\la\sigma_0\sigma_i\ra$. From this expression,
\beq
P(\sigma_0) = \frac12(1+m\sigma_0).
\eeq

\begin{figure}
\epsfig{file=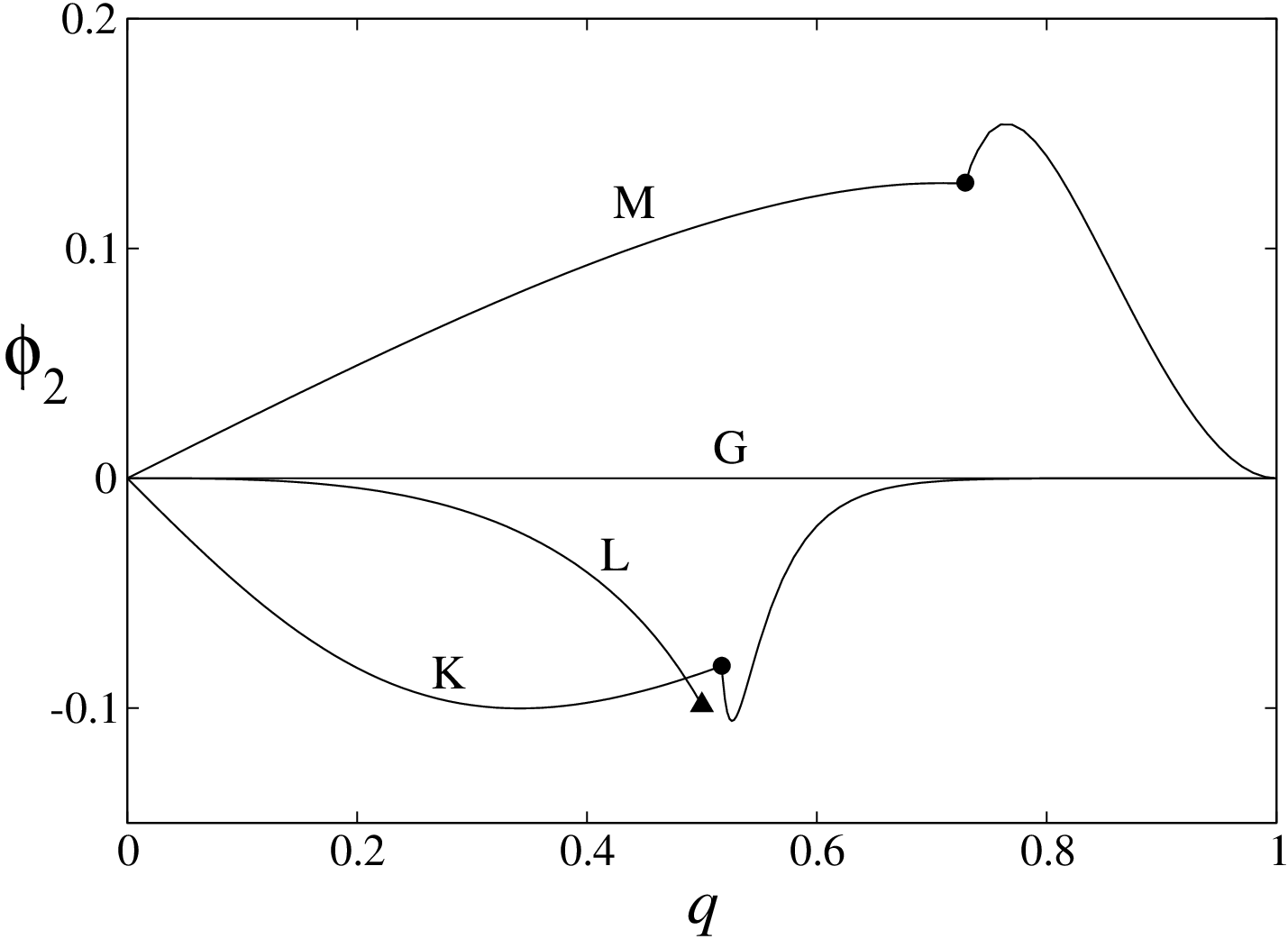,width=8.5cm}
\caption{Flux of energy $\phi_2$ from the second to the
first reservoir as a function of $q$ for the majority-vote
model (M), for the linear model (L), and for the model K.
For the Glauber model (G) the flux vanishes. The critical
points of models V and K occur at the kink of the curves
and are indicated by small full circles. The triangle 
represents the flux of energy for the voter model. The
results were obtained from the pair approximation on a
square lattice.}
\label{fluxpair}
\end{figure}

It is convenient to define the quantities
\beq
A_n = \frac{(m+r)^n}{(1+m)^{n-1}}, \qquad
B_n = \frac{(m-r)^n}{(1-m)^{n-1}}.
\eeq
In terms of these quantities we get
\beq
\la \sigma_0\sigma_1\ldots\sigma_n\ra
= \frac12(A_n - B_n),
\label{49a}
\eeq
\beq
\la \sigma_1\ldots\sigma_n\ra = \frac12(A_n + B_n).
\label{49b}
\eeq

To determine the time evolution of $m$, we multiply the
the master equation (\ref{45}) by $\sigma_0$ and sum
in the variables $\sigma_i$ after employing (\ref{46})
on the right-hand side of (\ref{45}). The result is
\beq
\frac{dm}{dt} = -m + 4a m + 2b (A_3+B_3).
\label{47}
\eeq
Similarly, we find the time evolution of $r$, which is
\beq
\frac{dr}{dt} = -2 r + 2a + 3(a+b)(A_2+B_2) + b (A_4+B_4).
\label{47a}
\eeq

\begin{figure}
\epsfig{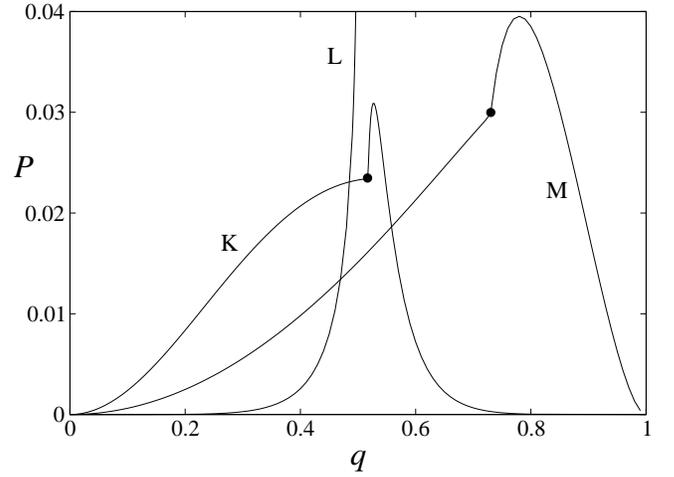}
\caption{Rate of entropy production ${\cal P}$
as a function of $q$ for the majority-vote
model (M), for the linear model (L), and for the model K.
For the linear model, ${\cal P}$ diverges at $q=1/2$.
The critical points of models V and K occur at the kink
of the curves and are indicated by small full circles.
The results were obtained from the fluxes 
of figure \ref{fluxpair} by the use of formula (\ref{51}).}
\label{prodpair}
\end{figure}

A stationary solution of these equations is $m=0$ and
$r$ the root of the equation
\beq
r = a + 3(a+b)r^2 + b r^4,
\label{48a}
\eeq
and corresponds to the disordered state. 
A stability analysis is obtained by the expansion 
of the right-hand side of (\ref{47}) in powers of
$m$. The solution becomes unstable when the coefficient
of $m$ vanishes, that is, when
\beq
4a + 12br^2 - 8br^3 = 1.
\label{48b}
\eeq
By inspection we see that the solution of
equations (\ref{48a}) and (\ref{48b}) is $r=1/3$
and $27a+7b=27/4$, or
\beq
17p + 20 q = 27,
\eeq
which describes the critical line in the plane 
$p,q$ that separates the disordered and ordered states.
The critical line ends on the voter point $p=1$ and
$q=1/2$. The critical point of the majority vote
model occurs when $p=q=27/37$, the
Glauber model when $q=3/5$ and $p=15/17$,
and the model K when $q=0.517246$ and $p=0.979710$.

Another solution corresponds to the ordered state
for which $m\neq0$. The values of $m$ and $r$
for this solution can be obtained by solving the
equations (\ref{47}) and (\ref{47a}), what we 
did numerically.

To determine the energy fluxes we observe that
they can be written in terms of correlations of
two and for sites. Using (\ref{36a}) and (\ref{36b}),
we may write $g_1$ and $g_2$, given by (\ref{32b})
and (\ref{32a}), as
\beq
g_1 = \frac12(\alpha_1-\sigma_0 \xi_1 q),
\label{32d}
\eeq
\beq
g_2 = \frac12(\alpha_2-\sigma_0 \xi_2 p).
\label{32c}
\eeq
We then replace these results in
\beq
\phi = \frac12(\phi_2-\phi_1) =\la \sigma_0 s (g_2-g_1)\ra.
\label{40}
\eeq
to find 
\beq
\phi = \frac12\la 
[\sigma_0 ( 4\xi_1 - 2\xi_2) - s(\xi_2 p - \xi_1 q)]\ra,
\label{40a}
\eeq
where we used the relations $\alpha_1 s = 2 \xi_1$
and $\alpha_2 s = 4 \xi_2$. Replacing the expression
of $\xi_1$ and $\xi_2$ given by equations (\ref{50b})
and (\ref{50a}) in this equation we find
\beq
\phi = \frac12\la 
\sigma_0 s \sigma_1\sigma_2\sigma_3\sigma_4
- s^2 (b  + a\sigma_1\sigma_2\sigma_3\sigma_4)\ra.
\label{40b}
\eeq

It is then straightforward to write $\phi$ in the form
\[
\phi = 2\la \sigma_0\sigma_1\sigma_2\sigma_3\ra - 2b
-2a \la\sigma_1\sigma_2\sigma_3\sigma_4\ra
\]
\beq
- 4(a+b)\la \sigma_1\sigma_2\ra
- 2(a+b)\la\sigma_1\sigma_3\ra,
\label{44}
\eeq
where the sites 1 and 3, and 2 and 4 are opposite 
in relation to the central site 0. This is accomplished
by using symmetric operations that leave a square 
lattice invariant.

From the stationary solutions for $m$ and $r$, 
calculated numerically, we have determined the values
of $\la \sigma_0\sigma_1\sigma_2\sigma_3\ra$,
$\la\sigma_1\sigma_2\sigma_3\sigma_4\ra$,
$\la \sigma_1\sigma_2\ra$, and 
$\la\sigma_1\sigma_3\ra$ by the use of (\ref{49a})
and (\ref{49b}). From these values we have
determined $\phi$ by (\ref{44}). In the stationary
state $\phi_1=-\phi_2$ so that $\phi_2=\phi$.
The flux of energy $\phi_2$ is shown in figure
\ref{fluxpair} as a function of $q$ for
the majority model, the model K and for the
linear model. From $\phi_2$, we have determined
the rate of entropy production by (\ref{51}),
which is shown in figure \ref{prodpair}
as a function of $q$ for
the majority model, the model K and for the
linear model.

For the Glauber model, our numerical calculations obtained
from the pair approximation show that the flux $\phi_2$
vanishes for any value of $q$. This result shows that this
approximation is capable of preserving the detailed balance
condition for the models that hold this property as happens
with the Glauber model.

\section{Conclusion}

We have shown that the models of opinion dynamics can
be framed within the stochastic thermodynamics
from which we may construct a nonequilibrium
thermodynamics of opinion dynamics. To this end
we postulated an energy function, which we have
called the opinion function, which is absent
in the original definition of a model as it is
defined by the transition rate. From the
energy function, we define by the use of formula
(\ref{12}) the various components of the transition
rate that defines the model that we wish to study.
Each component is understood as describing  
the contact with heat reservoirs at distinct temperatures.
As the temperatures are different from one another,
the system will be in a nonequilibrium state at
the stationary state, which is one of the main feature
of the opinion dynamics. 

Given a model, we face the problem of finding 
the energy function. Here, we have circumvented
this problem by postulating a certain energy function
and then determining the possible transition rates
that follows from that energy function. In the present
case we adopted the Ising energy function 
with nearest neighbor interactions, given by (\ref{30})
and then proved that it leads to the generic-vote model
by using (\ref{23}).

The idea of using heat reservoirs at distinct temperatures
or effective temperatures to define transition rates
leading to nonequilibrium steady states is not new and
has been used before, for instance, in the analysis
of the generic-vote model \cite{drouffe1999}.
The reference to heat reservoirs or to effective
temperatures is usually nominal. Here we justify this
idea by presenting a systematic approach in which a given
state of the system is associated to just one of the heat
reservoirs, with a transition rate describing the contact
with the heat reservoir being given by (\ref{12}).
Our approach leads to the significant Clausius relation
(\ref{25}) between flux of entropy, flux of heat and
temperature of the reservoir, that characterizes 
thermodynamically the contact of a system with
a heat reservoir. 

We have applied the approach developed here to
nonequilibrium lattice models with two states that
include the majority-vote model in the square lattice
and the cubic lattice. In the square lattice it suffices
to use two heat reservoirs but in the cubic lattice it
is necessary three heat reservoirs to set up the transition
rates. We have determined the fluxes of energy from each
reservoir from which we determined the rate of entropy flux.
The critical phase transition from a disordered to an
ordered state that takes place in the models analyzed here
is reflected in the fluxes and the production of entropy as
a singularity occurring at the inflexion point of these
quantities. We have also used a pair approximation in which
case the singularity is characterized by a kink.

The approach proposed here is general and can be 
applied to stochastic models as long as their
dynamic rules can be expressed by transition rates 
that can be decomposed into transition rates 
describing each one of them the contact with
a heat reservoir.


\end{document}